\documentclass[aps,twocolumn,superscriptaddress]{revtex4}
\usepackage{epsfig}
\usepackage{times}
\usepackage{color}

\begin{document}

\title{If players are sparse social dilemmas are too:\\Importance of percolation for evolution of cooperation}

\author{Zhen Wang}
\affiliation{School of Physics, Nankai University, Tianjin 300071, China}
\affiliation{Department of Physics, Hong Kong Baptist University, Kowloon Tong, Hong Kong}

\author{Attila Szolnoki}
\affiliation{Institute of Technical Physics and Materials Science, Research Centre for Natural Sciences, Hungarian Academy of Sciences, P.O. Box 49, H-1525 Budapest, Hungary}

\author{Matja{\v z} Perc}
\email{matjaz.perc@uni-mb.si}
\affiliation{Faculty of Natural Sciences and Mathematics, University of Maribor, Koro{\v s}ka cesta 160, SI-2000 Maribor, Slovenia}

\begin{abstract}
Spatial reciprocity is a well known tour de force of cooperation promotion. A thorough understanding of the effects of different population densities is therefore crucial. Here we study the evolution of cooperation in social dilemmas on different interaction graphs with a certain fraction of vacant nodes. We find that sparsity may favor the resolution of social dilemmas, especially if the population density is close to the percolation threshold of the underlying graph. Regardless of the type of the governing social dilemma as well as particularities of the interaction graph, we show that under pairwise imitation the percolation threshold is a universal indicator of how dense the occupancy ought to be for cooperation to be optimally promoted. We also demonstrate that myopic updating, due to the lack of efficient spread of information via imitation, renders the reported mechanism dysfunctional, which in turn further strengthens its foundations.
\end{abstract}

\maketitle

Since the seminal paper on games and spatial chaos \cite{nowak_n92b}, spatial reciprocity has been built upon as a powerful mechanism for the promotion of cooperation \cite{nowak_11}. Alongside kin and group selection \cite{maynard_n64, hamilton_wd_jtb64a} as well as direct and indirect reciprocity \cite{nowak_n98, fort_pa03, panchanathan_n04, pacheco_ploscb06, ohtsuki_jtb06b}, it completes the list of the big five \cite{nowak_s06} held responsible for why we tend to overcome our selfishness for the greater common good. Aiding its popularity is certainly the fact that its workings can be described in a couple of lines. If the interactions amongst players are restricted to only a few individuals by means of a graph, then cooperators can survive by means of forming compact clusters, which minimizes the potential exploitation by defectors and protects those that are located in the interior of such clusters against an invasion. It is along the lines of this observation that studies on the evolution of cooperation have received a substantial boost, as evidenced in several reviews that capture succinctly recent advances on this topic \cite{doebeli_el05, szabo_pr07, roca_plr09, perc_bs10}.

One of the most notable spinoff discoveries stemming from the early works on the importance of spatial structure \cite{nowak_ijbc94, lindgren_pd94, nakamaru_jtb97, szabo_pr07} has been that complex networks, having the connectivity structure similar to that of social networks, are very beneficial for the evolution of cooperation \cite{abramson_pre01, santos_prl05, tomassini_ijmpc07, du_f_dga11, santos_pnas06, tang_epjb06, lozano_ploso08, gomez-gardenes_jtb08, poncela_njp07, kuperman_epjb08, wang_s_ploso08, fu_jtb10, gomez-gardenes_prl07, du_wb_epl09, pacheco_ploscb09, tomassini_g10}. More generally, it was discovered that the heterogeneity or diversity allows for cooperative behavior to prevail even if the temptations to defect are large \cite{perc_pre08, santos_n08, fu_pre08, perc_njp11, santos_jtb11b}. Recently, evolutionary games have also been studied in growing populations \cite{poncela_njp09, poncela_ploso08} and hierarchical structures \cite{lee_s_prl11}, thus elegantly continuing this line of research.

Another important avenue of research having its roots firmly in spatial games is the study of disordered environments \cite{vainstein_pre01}, which subsequently gave rise to studies clarifying the role of mobility in different evolutionary settings \cite{vainstein_jtb07, sicardi_jtb09, meloni_pre09, helbing_pnas09, droz_epjb09, cheng_hy_njp10}. It is by now a fact that mobility of players can pave the way towards a successful evolution of cooperation, even if the conditions are noisy and do not necessarily favor the spreading of cooperators. Apart from an early work on diluted lattices \cite{vainstein_pre01}, however, the primary impact of population density has not been explored. Given that the early experiments on the behavior of rats under crowded conditions revealed that too high population densities may induce a variety of destructive conditions, ranging from infant cannibalism over excessive aggression to increased mortality at all ages \cite{calhoun_sa62}, and that it was later confirmed that similar effects of overcrowding can be observed not just by rodents, but also by primates \cite{judge_ab97} and humans \cite{galle_s72}, we are therefore motivated to examine in detail the role of the population density by the resolution of social dilemmas. For this purpose, we study the evolution of cooperation in the prisoner's dilemma, the snowdrift game and the stag-hunt game on different lattices. All the simulation details are described in the Methods, while here we proceed with presenting the results.

\section*{Results}

\begin{figure}
\begin{center}
\includegraphics[width = 8cm]{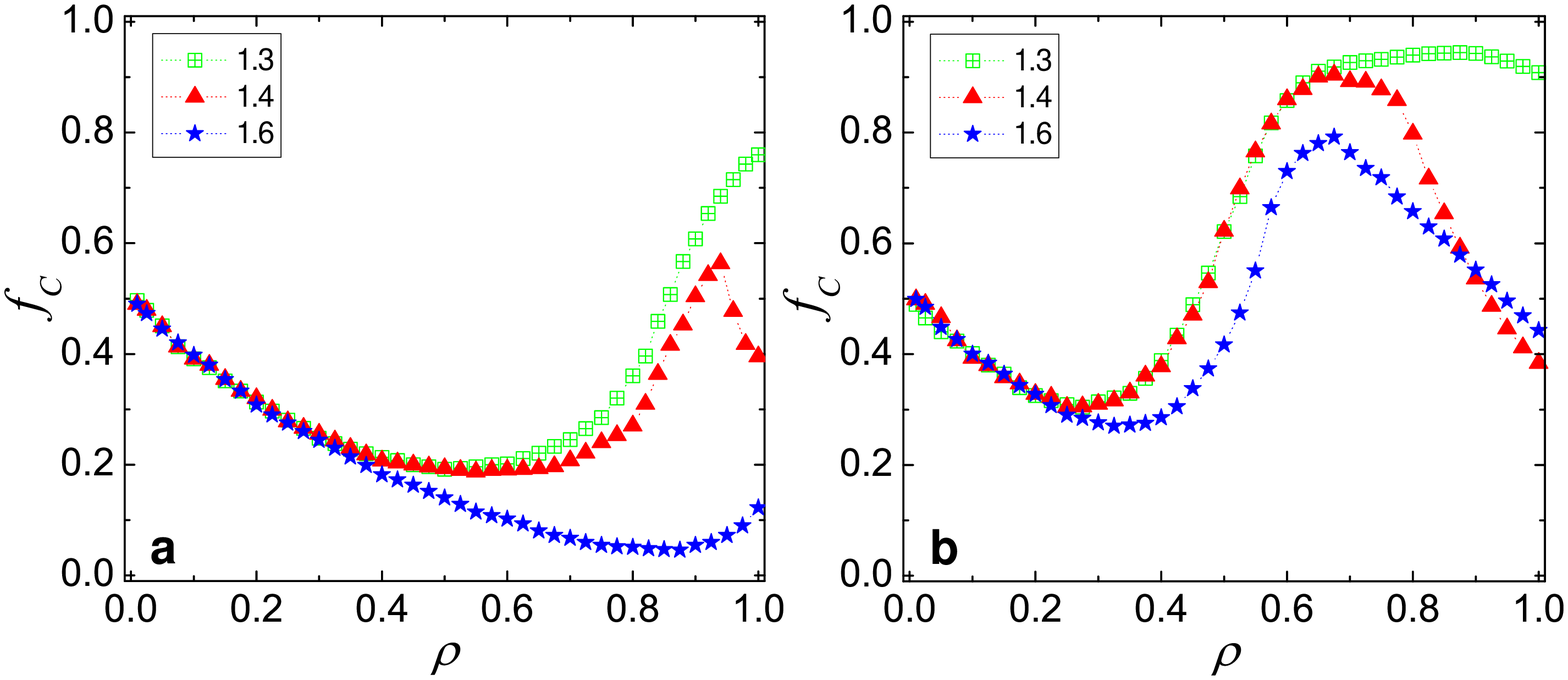}
\caption{\label{motivate}Fraction of cooperators $f_C$ in dependence on the population density $\rho$ for different values of the temptation to defect $b$ (see legend), as obtained for the prisoner's dilemma game on the square lattice under sequential updating. Results in panel (a) were obtained with the ``choosing the best'' strategy updating rule following \cite{vainstein_pre01}, while results in panel (b) were obtained by means of a stochastic version of the same rule. If players are no longer forced to strictly adopt the strategy of their best neighbor, the $f_{C}(\rho)$ dependence changes dramatically, exhibiting a consistent optimum at an intermediate value of $\rho$. This is because frozen states that do not correspond to the global optimum, as well as the sensitivity on initial conditions, especially at low $\rho$ values, are avoided. Introducing some uncertainty to strategy adoptions [panel (b)] thus helps to reveal a more interesting impact of population density as was previously reported to exist.}
\end{center}
\end{figure}

As motivational results presented in Fig.~\ref{motivate}(a) demonstrate, the usage of ``choosing the best'' strategy updating rule, during which a player follows unconditionally the strategy of its neighbor that has the largest payoff, has a detrimental impact on the outcome of social dilemmas on diluted lattices. Due to its deterministic nature, strategy updating by choosing the best player in the neighborhood frequently leads to the system being trapped into a frozen state, which however, does not correspond to its global optimum. Moreover, the final state depends sensitively on the initial configuration \cite{vainstein_pre01, arapaki_pa09}, especially at lower densities of players (low values of $\rho$). To avoid unwanted properties of deterministic strategy updating, we employ a stochastic updating rule with a direct noise parameter that allows ``irrational'' behavior, albeit with a small probability. This leads to qualitatively different results that become independent of the initial state if $\rho$ exceeds $0.2$. As Fig.~\ref{motivate}(b) highlights, there are intermediate population densities that play a more prominent and consistent role. Motivated by these results, we proceed with using the pairwise stochastic imitation $w(s_x \to s_y)$ (see Eq.~\ref{fermi}) introduced in the Methods. Our goal is to explore how the cooperation level depends on $\rho$, and how robust the outcomes are if using different host latices. As described in the Methods, a direct comparison is possible by means of normalizing $K$ (the uncertainty by strategy adoptions) with $k$ (the degree that characterizes different lattices).

\begin{figure}
\begin{center}
\includegraphics[width = 8.5cm]{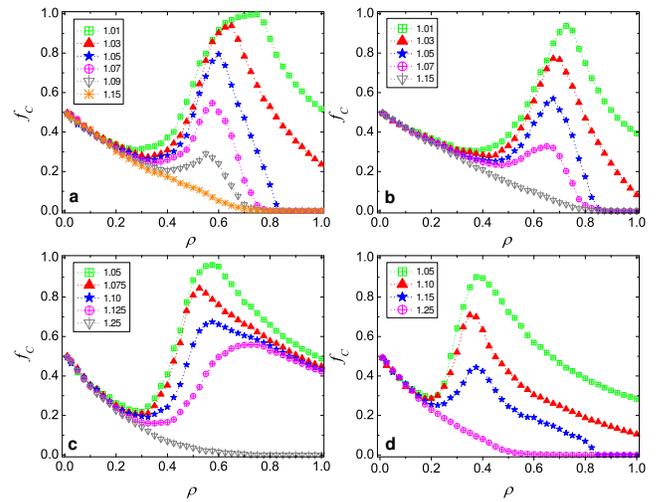}
\caption{\label{pd}Fraction of cooperators $f_C$ in dependence on the population density $\rho$ for different values of the temptation to defect $b$ (see legend), as obtained for the prisoner's dilemma game on the square lattice [panel (a)], the honeycomb lattice [(panel (b)], the triangular lattice [(panel (c)] and the cubic lattice [(panel (d)]. It can be observed that, regardless of the type of the underlying interaction graph, there always exists and intermediate value of $\rho$ at which $f_C$ is maximal. More importantly, if $b$ is close to the critical value at which cooperators would normally die out, the optimal population density is strongly related to the percolation threshold of the interaction graph. For the square lattice the latter is $\pi=0.5$, for the honeycomb lattice it is $\pi=0.6527$, for the triangular lattice it is $\pi=0.3472$, while for the cubic lattice it is $\pi=0.2488$. Accordingly, percolation plays a key role by the resolution of social dilemmas by means of drastically elevating the effectiveness of spatial reciprocity. Error bars are comparable to the size of symbols.}
\end{center}
\end{figure}

Since it represents the hardest social dilemma to solve, we stay with the focus on the prisoner's dilemma game, and present in Fig.~\ref{pd} its outcome on four different lattices in dependence on the population density $\rho$. It can be observed that there always exists an intermediate value of $\rho$ at which the fraction of cooperators $f_C$ is maximal. Depending on the temptation to defect $b$ and the type of the interaction graph, however, the maxima of $f_C$ occur at different $\rho$. A closer inspection reveals that in fact the shifts are strongly related to the percolation thresholds \cite{stauffer_94} of the underlying lattices. Accordingly, in panel (b) the maxima occur at the highest values of $\rho$, as the honeycomb lattice indeed has the highest percolation threshold ($\pi=0.6527$) amongst the four considered lattices. Conversely, the cubic lattice [see panel (d)], having $\pi=0.2488$, requires the lowest density (largest sparsity) of players for the evolution of cooperation to be optimally promoted.

Apart from the graph-specific dependence of the optimal $\rho$, there are also general features common to all four interactions graphs. In the $\rho \to 0$ limit the majority of players will have no neighbors, and hence $f_C$ simply mirrors back the initial state that is $\rho=0.5$. As $\rho$ increases, the few existing links between players enable defectors to exploit cooperators without having to fear the consequences of spatial reciprocity. Note that for sufficiently small $\rho$ many players, as well as large portions of the graph as a whole, will still be disconnected, hence prohibiting cooperators to form compact clusters and utilizing spatial reciprocity to protect themselves against invading defectors. Because of the random initial state, the initial invasion of defectors will always be successful, regardless of the value of $b$. But further invasions are subsequently hindered by the lack of connections between players that are utilizing different strategies, and hence at low values of $\rho$ the decay of $f_C$ is universal for all values of $b$. For larger $\rho$, however, the outcome becomes independent of the initial state and the temptation to defect more and more crucial. For higher values of $b$ the $f_C$ trend simply continues downward as $\rho$ increases, which indicates that new cooperative players simply serve as ``sitting ducks'' for defectors. At lower values of $b$ cooperators are able to utilize the enhanced interconnectedness between them to form compact clusters, while at the same time benefiting from the dilution that prohibits defectors to exploit them with the same efficiency as on a fully populated graph. Accordingly, $f_C$ peaks at an intermediate (optimal) value of $\rho$, which is a bit higher but close to the percolation threshold of the underlying interaction graph \cite{stauffer_94}. The fact that it is a bit higher is simply a consequence of the fact that not all players will be cooperators, and hence for cooperation to start percolating the fraction needs to be somewhat higher to offset the defectors.

\begin{figure}
\begin{center}
\includegraphics[width = 8cm]{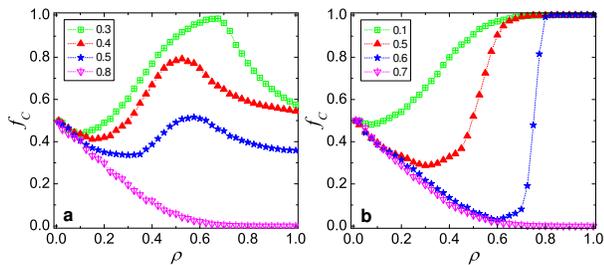}
\caption{\label{sdsh}Fraction of cooperators $f_C$ in dependence on the population density $\rho$ for different values of the cost-to-benefit ration $r$ (see legend), as obtained on the square lattice for the snowdrift [panel (a)] and the stag-hunt [(panel (b)] game by using strategy imitation defined by Eq.~\ref{fermi}. For the snowdrift game the results are qualitatively identical as for the prisoner's dilemma game in that there exists an intermediate value of $\rho$ where $f_C$ is maximal. Since the stag-hung game is a less severe social dilemma, the exceeding of the percolation threshold is sufficient for eliciting the all-$C$ state. Regardless of the governing social dilemma, however, the percolation threshold is an important benchmark for how high a population density ought to be for cooperation to thrive. Error bars are comparable to the size of symbols.}
\end{center}
\end{figure}

Results presented in Fig.~\ref{sdsh} for the snowdrift [panel (a)] and the stag-hunt game [panel (b)] further add to the general validity of the outlined mechanism. The percolation threshold still marks the advent of enhanced cooperation, although for the stag-hunt game [panel (b)], which is in itself more lenient for the evolution of cooperation, the all-$C$ state rather than an optimum in $f_C$ sets in. Along with the results reported previously for the multi-player interaction public goods game \cite{wang_z_pre12}, this leads us to the conclusion that a population density close to the percolation threshold is optimal for the successful evolution of cooperation. In particular, the players are connected enough to transfer the more advantageous mutually beneficial cooperative strategy, while simultaneously the graph is diluted enough for the defectors to be unable to invade cooperators effectively. Crucial for this scenario to be valid is thus percolation, and directly related to that the fact that information can spread efficiently by means of stochastic strategy imitation.

\begin{figure}
\begin{center}
\includegraphics[width = 8cm]{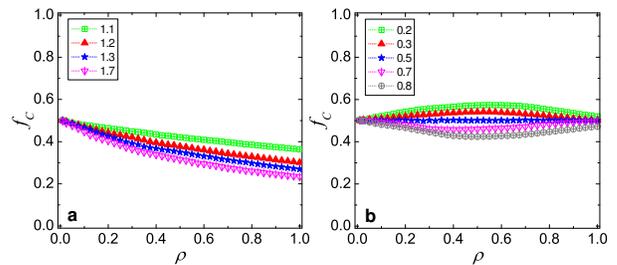}
\caption{\label{pdsdmyop}Fraction of cooperators $f_C$ in dependence on the population density $\rho$, as obtained on the square lattice under myopic updating (see Eq.~\ref{myop}), for the prisoner's dilemma [panel (a)] and the snowdrift game [panel (b)]. Regardless of the governing temptation to defect $b$, in the prisoner's dilemma game the population density has a monotonous impact on $f_C$. In the snowdrift game, however, the myopic updating can lead to a role-separating distribution of $C$s and $D$s that is reminiscent of anti-ferromagnetic order. Yet the increase of cooperation stemming from this is practically negligible, especially if compared to the results presented in Fig.~\ref{sdsh}(a), where imitation was used as the driving force behind the evolution of strategies. These observations confirm that the mechanism by means of which the percolation threshold is established as the optimal population density for the resolution of social dilemmas relies on the percolation of cooperators and the directly related effective spread of information via strategy imitation. Myopic updating hinders the later, and hence the mechanism becomes dysfunctional. Error bars are comparable to the size of symbols.}
\end{center}
\end{figure}

The validity of this argument can be tested effectively by replacing the strategy updating via imitation by the so-called myopic strategy updating rule \cite{sysiaho_epjb05, szabo_pre10}. In this case, every player makes decisions locally as an individual, always assuming an unchanged neighborhood (see Methods for details). If the more successful strategy is not adopted, the existence or absence of the percolation of players becomes an uncritical property of the interaction topology, and hence it is expected that the outlined mechanism will no longer work. Results presented in Fig.~\ref{pdsdmyop} fully confirm this expectation, as indeed neither for the prisoner's dilemma [panel (a)] nor for the snowdrift game [panel (b)] a decisive importance of an intermediate population density cannot be observed. As the population density increases, the fraction of cooperators decays more or less fast in the prisoner's dilemma game, indicating that the new connections amongst players mainly serve the defecting strategy by allowing an ever increasing efficiency of local exploitation. The situation for the snowdrift game, shown in Fig.~\ref{pdsdmyop}(b), is a bit different because the myopic strategy updating allows for the emergence of a role-separating distribution of $C$s and $D$s independently on the value of $r$, which is reminiscent of anti-ferromagnetic order \cite{szabo_pre10}. The increase in the level of cooperation, however, is significantly lower than reported in Fig.~\ref{sdsh}(a) for imitation.

\begin{figure}
\begin{center}
\includegraphics[width = 8.2cm]{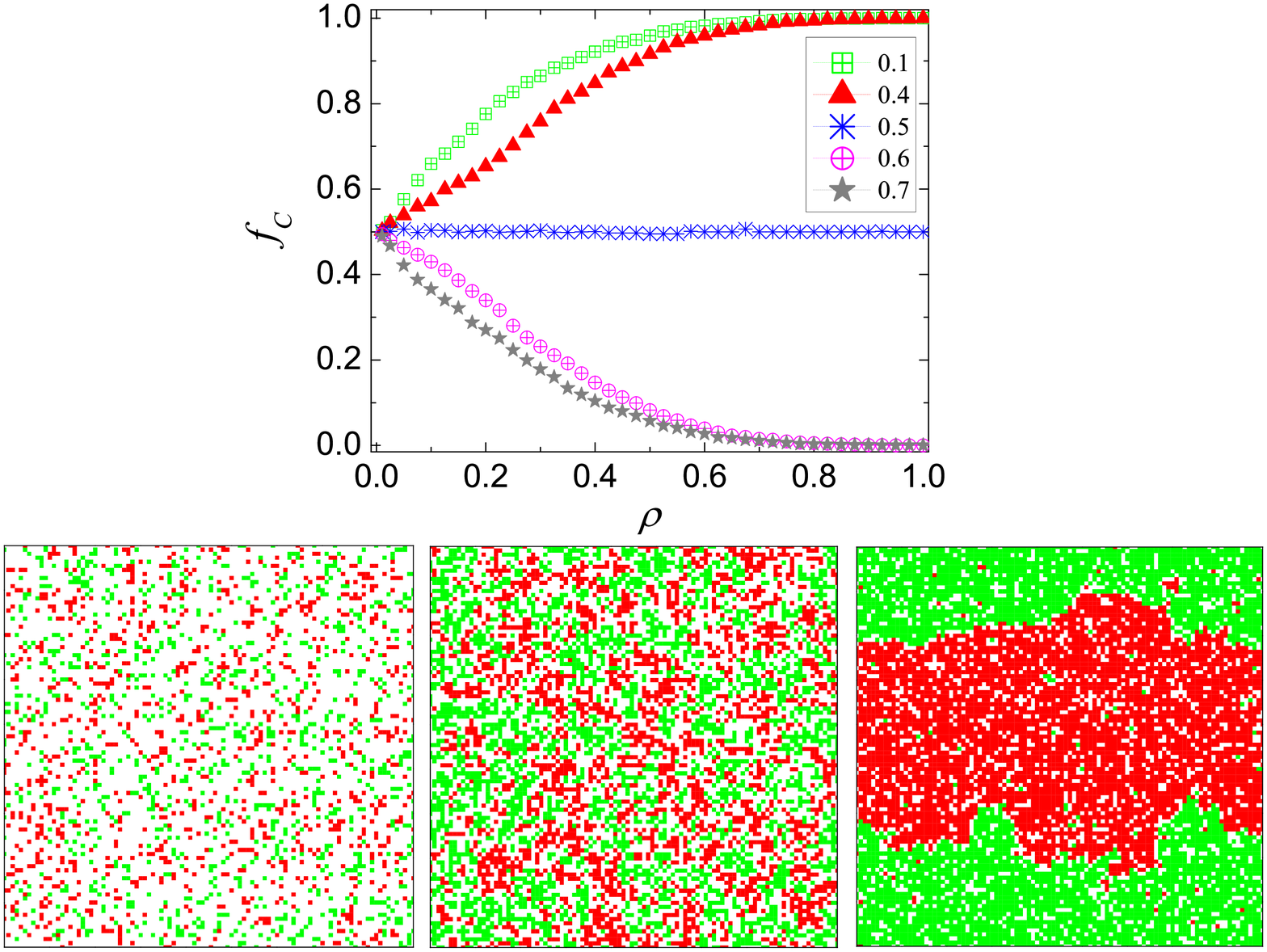}
\caption{\label{shmyop}Fraction of cooperators $f_C$ in dependence on the population density $\rho$, as obtained on the square lattice under myopic updating (see Eq.~\ref{myop}), for the stag-hunt game. As reported for the prisoner's dilemma and the snowdrift game in Fig.~\ref{pdsdmyop}, here too the population density has a monotonous impact, which however depends on the value of the cost-to-benefit ratio $r$. If $r<0.5$ the final destination is the all-$C$ phase, and accordingly, increasing $\rho$ leads progressively towards this solution. Conversely, for $r>0.5$ the final outcome on the fully populated lattice is the all-$D$ phase, and thus as $\rho$ increases $f_C$ decays. At $r=0.5$, however, there is a transition from the pure $C$ to the pure $D$ phase, and in fact on a fully populated lattice both are equally probable, hence $f_C=0.5$. For low values of $\rho$, however, the manifestation of $f_C=0.5$ is not by means of an eventual evolution of either a pure $C$ or a pure $D$ phase, but rather by the simultaneous yet isolated coexistence of both phases, as demonstrated by the characteristic snapshots in the bottom row left ($\rho=0.2$) and middle ($\rho=0.5$). If $\rho$ is sufficiently large, however, the original solution is recovered. The right snapshot was obtained at $\rho=0.8$ and demonstrates such a scenario, but the stationary state (which will be either a pure $C$ or a pure $D$ phase) is not yet reached. In the snapshots (bottom row) white denotes vacant sites, while green and red are cooperators and defectors, respectively.}
\end{center}
\end{figure}

The outcomes of the stag-hunt game under myopic updating presented in Fig.~\ref{shmyop} also agree with our expectations, only that in this case $r=0.5$ constitutes a transition point, above (below) which a pure $D$ ($C$) phase evolves. Accordingly, increasing $\rho$ towards one simply drives the system towards the expected state. The percolation threshold plays not role at all. Exactly at $r=0.5$, however, both the pure $C$ and the pure $D$ phase are equally probable. At low population densities (left and middle snapshot) both phases can coexist isolated from one another, hence yielding $f_C=0.5$, while at sufficiently high populations densities (right snapshot) a pure phase will eventually be reached (not shown) but since both outcomes are equally probable $f_C$ is again $0.5$. Regardless of the studied social dilemma, and also regardless of the type of the interaction graph, myopic updating cannot support an efficient transfer of information between the players, and thus renders the reaching of the percolation threshold with the population density irrelevant. This in turn confirms the validity of our arguments and establishes the percolation threshold as the key property of a graph that determine the optimal population density.

\section*{Discussion}
Previous studies highlighted that imitation plays a decisive role by the evolution of strategies amongst humans. By building on this fact, we have shown that the percolation threshold of the matrix that determines the interactions between players constitutes the optimal population density for the resolution of social dilemmas that are governed by pairwise interactions. For the mechanism to work, some level of uncertainty by strategy adoptions is crucial as it prevents the system being trapped into a frozen state, and it alleviates the dependence on initial conditions, especially if the population density is high. We have demonstrated that the results are valid for all social dilemma games and on a wide class of different lattices, which together with the previous results on the public goods game that is governed by group interactions \cite{wang_z_pre12}, firmly solidifies the percolation threshold as the crucial property that determines the optimal population density for the evolution of cooperation. As a reverse test, we have verified the validity of our arguments by means of the myopic updating rule, under which players are no longer able to exchange information directly between each other. Expectedly, we have found that the percolation threshold no longer
has a decisive impact on the outcome of the three considered social dilemmas. This confirms that the percolation threshold constitutes the optimal population density for the resolution of social dilemmas by ensuring the percolation of cooperators and the directly related effective spread of information via strategy imitation.

It can be argued that the optimal population density amplifies the mechanism of spatial reciprocity \cite{nowak_n92b}. If the population density is too low, vacant sites prohibit the formation of compact clusters by cutting short the communication paths between the cooperators. Too high populations densities, on the other hand, enable an effective invasion of defectors, which again disrupts reciprocity amongst cooperators by splitting them up into isolated clusters. Presented results thus allow us to understand the impact of population density on the resolution of social dilemmas through the concept of percolation, and by doing so they provide an interesting interdisciplinary link between statistical physics and the evolution of cooperation.

\section*{Methods}
Within this work we consider the spatial prisoner's dilemma, the spatial snowdrift and the spatial stag-hunt game. In all three games players can choose either to cooperate ($s_x=C=1$) or to defect ($s_x=D=0$), whereby mutual cooperation yields the reward $R$, mutual defection leads to punishment $P$, and the mixed choice gives the cooperator the sucker's payoff $S$ and the defector the temptation $T$. Depending on the rank of these four payoffs we have the prisoner's dilemma game if $T>R>P>S$, the snowdrift game if $T>R>S>P$, and the stag-hunt game if $R>T>P>S$. For simplicity, we here take $R=1$ and $P=0$, which imposes boundaries on the remaining two payoffs of the form $-1 \leq S \leq 1$ and $0 \leq T \leq 2$. Further zooming in on the most relevant features of the three dilemmas, we take for the prisoner's dilemma the temptation to defect $T=b$ and the punishment for mutual defection $P=0$ \cite{nowak_n92b}, for the snowdrift game we take $T=1+r$ and $S=1-r$ \cite{szolnoki_epl09}, while for the stag-hunt game we use $T=r$ and $S=-r$, where $r$ in both cases is the cost-to-benefit ratio. As interaction graphs that characterize the topology of the matrix containing players, we employ the square, honeycomb, triangular and the cubic lattice, each with linear size $L$ and only a fraction $\rho$ of occupied nodes. The remaining $1-\rho$ nodes are left vacant. The random dilution is performed only once at the start of the game.

Following the initialization, we carry out Monte Carlo simulations comprising the following elementary steps. First, a randomly selected player $x$ acquires its payoff $p_x$ by playing the game with its $k$ neighbors, as specified by the underlying interaction graph. Next, one randomly chosen neighbor, denoted by $y$, also acquires its payoff $p_y$ by playing the game with its four neighbors. Lastly, player $x$ tries to enforce its strategy $s_x$ on player $y$ in accordance with the probability
\begin{equation}
w(s_x \to s_y)=\frac{1}{1+\exp[(p_{y}-p_{x})/(kK)]}\ \,
\label{fermi}
\end{equation}
where $K$ determines the level of uncertainty by strategy adoptions \cite{szabo_pr07}, which can be attributed to errors in judgment due to mistakes and external influences that affect the evaluation of the opponent. Without loss of generality we set $K=0.1$ normalized with the degree of the underlying lattice $k$, implying that better performing players are readily imitated, but it is not impossible to adopt the strategy of a player performing worse. Each Monte Carlo step (MCS) gives a chance for every player to enforce its strategy onto one of the neighbors (if they exist, which at sufficiently small $\rho$ will not be the case) once on average. The average density of cooperators $f_{C}=\rho^{-1} L^{-2}\sum_x s_x$ is determined in the stationary state after sufficiently long relaxation times. Depending on the actual conditions the linear system size was varied from $L=200$ to $1200$ and the relaxation time was varied from $10^4$ to $10^6$ MCS to ensure proper accuracy. The presented results are independent of the system size and valid in the large size limit.

As an alternative to imitation we also consider the myopic updating rule, where instead of comparing payoffs with a neighboring player and determining $w(s_x \to s_y)$ as the probability of strategy transfer (see Eq.~\ref{fermi}), a randomly chosen player $x$ changes its strategy $s_x$ to the other strategy $s_x^{\prime}$ with a probability
\begin{equation}
q(s_x^{\prime} \to s_x) =\frac{1}{1+\exp[(p_x-p_x^{\prime})/(kK)]}\ \,
\label{myop}
\end{equation}
where $p_x$ and $p_x^{\prime}$ are the payoffs of player $x$ when playing $s_x$ and $s_x^{\prime}$ in its neighborhood. The simulation details and the determination of $f_{C}$, however, are the same as by imitation.

\noindent \\ \\ \textbf{Acknowledgments} \\
This research was supported by the Hungarian National Research Fund (Grant K-73449) and the Slovenian Research Agency (Grant J1-4055).

\noindent \\ \textbf{Author contributions} \\
Zhen Wang, Attila Szolnoki and Matja{\v z} Perc designed and performed the research as well as wrote the paper.

\noindent \\ \textbf{Competing financial interests} \\
The authors declare no competing financial interests.

\end{document}